\documentclass[aps,prd,twocolumn,superscriptaddress,preprintnumbers,floatfix,nofootinbib,showpacs]{revtex4}

\usepackage{color}
\usepackage[dvips]{graphicx}

\newcommand{\lsim}{\mathrel{\mathop{\kern 0pt \rlap {\raise.2ex\hbox{$<$}}}\lower.9ex\hbox{\kern-.190em $
\sim$}}}

\newcommand{\gsim}{\mathrel{\mathop{\kern 0pt \rlap{\raise.2ex\hbox{$>$}}}\lower.9ex\hbox{\kern-.190em $\sim
$}}}
\newcommand{\beq}    {\begin{equation}}
\newcommand{\eeq}    {\end{equation}}

\newcommand{\be}{\begin{equation}}
\newcommand{\ee}{\end{equation}}
\newcommand{\beqarr}{\begin{eqnarray}}
\newcommand{\eeqarr}{\end{eqnarray}}

\newcommand{\vesc}{v_{\rm esc}}

\begin{document}

\title{Interpreting the recent results on direct search for dark matter particles
in terms of relic neutralinos}

\preprint{Preprint number: DFTT 12/2008}


\author{A. Bottino}
\affiliation{Dipartimento di Fisica Teorica, Universit\`a di Torino \\
Istituto Nazionale di Fisica Nucleare, Sezione di Torino \\
via P. Giuria 1, I--10125 Torino, Italy}
\author{F. Donato}
\affiliation{Dipartimento di Fisica Teorica, Universit\`a di Torino \\
Istituto Nazionale di Fisica Nucleare, Sezione di Torino \\
via P. Giuria 1, I--10125 Torino, Italy}
\author{N. Fornengo}
\affiliation{Dipartimento di Fisica Teorica, Universit\`a di Torino \\
Istituto Nazionale di Fisica Nucleare, Sezione di Torino \\
via P. Giuria 1, I--10125 Torino, Italy}
\author{S. Scopel}
\affiliation{Korea Institute for Advanced Study \\
Seoul 130-722, Korea}
\date{\today}

\begin{abstract}

The most recent results from direct searches for dark matter particles in the
galactic halo are examined in terms of an effective Minimal
Supersymmetric extension of the Standard Model at the
electroweak scale without gaugino masses unification. We show that
the annual modulation effect at  8.2 $\sigma$ C.L.
recently presented by the DAMA Collaboration, as the result of a
combined analysis of the DAMA/NaI and the DAMA/LIBRA experiments
for a total exposure of 0.82 ton yr, fits remarkably well  with what
expected for relic neutralinos for a wide variety of WIMP distribution functions.
Bounds derivable from other measurements of direct searches for dark matter particles
are analyzed. We stress the role
played by the uncertainties affecting the neutralino--quark couplings
arising from the involved hadronic quantities.
We also examine how present data on cosmic antiprotons can help
in constraining the neutralino configurations selected by the
DAMA effect, in connection with the values of the astrophysical parameters.
 Perspectives for measurement of antideuterons
possibly produced  in the galactic halo by  self--annihilation of
neutralinos belonging to the DAMA configurations are examined.
Finally, we discuss how findings at LHC would impact on these issues.

\end{abstract}


\maketitle

\section{Introduction}
\label{sec:intro}

We are at present witnessing a great activity in running and
 preparing experimental projects of
direct searches for dark matter (DM) particles in the galactic halo \cite{taup2007}.

A number of new results came out recently. Most notably, the DAMA
Collaboration, by analyzing the data obtained with the
DAMA/LIBRA experiment \cite{dama/libra}, has confirmed the evidence
for the annual modulation effect already measured with the previous
DAMA/NaI experiment \cite{dama/nai}.
As a result of the
combined analysis of the DAMA/NaI and the DAMA/LIBRA experiments,
for a total exposure of 0.82 ton yr,  the effect of the annual modulation
is now at 8.2 $\sigma$ C.L. \cite{dama/libra}.
Other Collaborations have reported upper bounds on the WIMP--nucleon
 scalar cross section
\cite{taup2007,xe,cdms,kims} (as usual here  WIMP  stands for a generic
Weakly Interacting
Massive Particle), using other approaches and scaling from different nuclei.

In this paper we discuss the relevance of these experimental
results for the most widely discussed candidate for cold dark
matter, the neutralino.
Neutralino configurations (concerning here mainly light neutralinos) which
fit the annual modulation data are extracted from the DAMA
results and confronted with the current measurements on galactic antiprotons.
Perspectives of seeing signals due to these configurations in forthcoming
measurements of galactic antideuterons are examined.
The current experimental upper bounds on the WIMP-nucleon scalar cross section
are also scrutinized.

In all our considerations we explicitly keep into account
 the uncertainties affecting the neutralino--quark couplings
because of the involved hadronic quantities, typically the pion--nucleon
sigma term $\sigma_{\pi N}$, the quantity $\sigma_0$ which is related to
the SU(3) symmetry breaking
and a quark mass ratio $r$ (see definitions later on).

In the analysis of the DAMA data, we discuss both the case
in which the channeling effect \cite{channeling} is
included in their analysis as the one where this effect is not taken
into account; these two cases are treated separately.
We recall that the channeling effect occurs when an ion, 
traversing a crystalline structure along a path (quasi-) parallel
to crystallographic axes or planes, has a deeper penetration than
in an amorphous material of the same composition \cite{drob}. 
In Ref. \cite{channeling} it is shown that the occurrence of the channeling 
effect makes the response of the DAMA NaI(Tl) detector more
sensitive to WIMP-nucleus interactions than previously estimated. 

Finally, in the present paper it is stressed that CERN LHC
(Large Hadron Collider), which hopefully will shed light on the
Higgs sector and supersymmetric extensions of the Standard Model (SM), 
has a remarkable discovery potential in terms of
the population of  light neutralinos which fit the annual modulation data.

The plan of the paper is the following. In Sect. II the main features of the supersymmetric
model employed in the present paper are delineated. Sect. III is devoted to a recollection of the
main phenomenological properties used in our analysis, and Sects. IV--V to a detailed comparison
of our theoretical evaluations to the current results of WIMP direct detection. In Sect. VI we discuss the
constraints implied by measurements of galactic antiprotons and in Sect. VII we examine the perspectives for
forthcoming measurements on cosmic antideuterons. Impact of LHC on these issues is discussed
in Sect. VIII, and finally conclusions are drawn in Sect. IX.

\section{The supersymmetric model}
\label{sec:susy}

The supersymmetric scheme we employ in the present paper is the
one described in Ref. \cite{lowneu}: an effective MSSM scheme
(effMSSM) at the electroweak scale, with the following independent
parameters: $M_1, M_2, \mu, \tan\beta, m_A, m_{\tilde q}, m_{\tilde l}$
and $A$. Notations are as
follows: $M_1$ and $M_2$  are the U(1) and  SU(2)  gaugino masses
(these parameters are taken here to be positive),
$\mu$ is the Higgs mixing mass parameter, $\tan\beta$ the
ratio of the two Higgs v.e.v.'s, $m_A$ the mass of the CP-odd
neutral Higgs boson, $m_{\tilde q}$ is a squark soft--mass common
to all squarks, $m_{\tilde l}$ is a slepton soft--mass common to
all sleptons, and $A$ is a common dimensionless trilinear parameter
for the third family, $A_{\tilde b} = A_{\tilde t} \equiv A
m_{\tilde q}$ and $A_{\tilde \tau} \equiv A m_{\tilde l}$ (the
trilinear parameters for the other families being set equal to
zero). In our model, no gaugino--mass unification at a Grand Unified (GUT)
scale is assumed. The lightest neutralino is required to be the lightest
supersymmetric particle and stable (because of R--parity conservation).

In the present paper the numerical analyses are performed by a
scanning of the supersymmetric parameter space, with the following
ranges of the
MSSM parameters: $1 \leq \tan \beta \leq 50$,
$100 \, {\rm GeV} \leq |\mu| \leq 1000 \, {\rm GeV},
5 \, {\rm GeV} \leq M_1 \leq 500 \, {\rm GeV},
100 \, {\rm GeV} \leq M_2 \leq 1000 \, {\rm GeV},
100 \, {\rm GeV} \leq m_{\tilde q}, m_{\tilde l} \leq 3000 \, {\rm GeV }$,
$90\, {\rm GeV }\leq m_A \leq 1000 \, {\rm GeV }$,
$-3 \leq A \leq 3$.

The following experimental constraints are imposed: accelerators data on
supersymmetric and Higgs boson searches (CERN $e^+ e^-$ collider LEP2
\cite{LEPb} and Collider Detectors D0 and CDF at Fermilab \cite{cdf});
measurements of the $b \rightarrow s + \gamma$ decay process \cite{bsgamma}:
2.89 $\leq B(b \rightarrow s + \gamma) \cdot 10^{-4} \leq$ 4.21 is employed
here: this interval is larger by 25\% with respect to the experimental
determination  \cite{bsgamma} in order to take into account theoretical
uncertainties in the SUSY contributions  \cite{bsgamma_theorySUSY} to the
branching ratio of the process (for the Standard Model calculation, we employ
the recent NNLO results from Ref.  \cite{bsgamma_theorySM}); the upper bound on
the branching ratio $BR(B_s^{0} \rightarrow \mu^{-} + \mu^{+})$ \cite{bsmumu}: we
take $BR(B_s^{0} \rightarrow \mu^{-} + \mu^{+}) < 1.2 \cdot 10^{-7}$;
measurements of the muon anomalous magnetic moment $a_\mu \equiv (g_{\mu} -
2)/2$: for the deviation $\Delta a_{\mu}$ of the  experimental world average
from the theoretical evaluation within the Standard Model we use here the range
$-98 \leq \Delta a_{\mu} \cdot 10^{11} \leq 565 $, derived from the latest
experimental  \cite{bennet} and theoretical \cite{bijnens} data.

Also included is the cosmological constraint that the neutralino
relic abundance does not exceed the maximal allowed value for cold dark matter,
{\it i.e.} $\Omega_{\chi} h^2 \leq (\Omega_{CDM} h^2)_{\rm max}$.
We set $(\Omega_{CDM} h^2)_{\rm max}  = 0.122$, as derived at a 2$\sigma$ level from
the results of Ref. \cite{wmap5}. We recall that this cosmological upper bound
implies  on the neutralino mass the lower limit $m_{\chi} \gsim$ 7 GeV \cite{lowneu}.

\section{Phenomenology related to the direct search of galactic WIMPs}
\label{sec:phen}

In case of WIMPs with coherent interactions with nuclei
 the differential event rate $dR/dE_R$ ($E_R$ being
the nuclear recoil energy) measured in WIMP direct searches
can be written as

\begin{equation}
\frac{dR}{dE_R} = N_T \frac{\rho_0}{m_\chi}\frac{m_N}{2 \mu_1^2}
A^2 \xi \sigma_{\rm scalar}^{(\rm nucleon)} F^2(E_R)\, {\cal I}(v_{\rm min})\, ,
\label{eq:rate}
\end{equation}
where:
\begin{equation}
{\cal I}(v_{\rm min}) = \int_{w \geq v_{\rm min}} d^3 w \;\;\frac{f_{\rm ES}(\vec w)}{w}\, .
\label{eq:ivmin}
\end{equation}
In  these formulae, notations are: $N_T$ is the number of target
nuclei per unit mass, $m_N$ is the nucleus
mass, $\mu_1$ is the WIMP--{\em nucleon} reduced mass, $A$ the nuclear
mass number, $\sigma_{\rm scalar}^{(\rm nucleon)}$ is the WIMP--nucleon coherent cross section,
$F(E_R)$ is the nuclear form factor, $\xi$ is the fraction of the mass
density of the WIMP in terms of the total local density for
non-baryonic dark matter $\rho_0$ ({\it i.e.}: $\xi = \rho_W/\rho_0$),
$f_{\rm ES}(\vec w)$ and $\vec w$ denote the velocity distribution
function (DF) and WIMP velocity in the Earth frame, respectively
($w=|\vec w|$), $v_{\rm min}$ is the minimal
Earth--frame WIMP velocity which contributes to a given recoil energy
$E_R$: $v_{\rm min} = [m_N E_R/(2 \mu_A^2)]^{1/2}$ ($\mu_A$ being
the WIMP--{\em nucleus} reduced mass).

 It is convenient to define a velocity distribution
function in the Galactic rest frame $f(\vec{v})$, where $\vec{v} =
\vec{w} + \vec{v}_{\oplus}$, $\vec{v}_{\oplus}$ being the Earth
velocity in the Galactic rest frame. The Earth frame velocity DF is
then obtained by means of the transformation: $f_{\rm ES}(\vec
w)=f(\vec w + \vec v_\oplus)$.
By definition, the velocity distribution function $f(\vec{v})$
is given by the six--dimensional phase--space distribution function
$F(\vec r, \vec v)$ evaluated at the Earth location ${\vec R}_0$ in the Galaxy, {\it i.e.}
$f(\vec{v}) = F({\vec R}_0, \vec v)$. The
velocity DF $f(\vec v)$ is truncated at a maximal escape velocity
$v_{\rm esc}$, since the gravitational field of the Galaxy cannot
bound arbitrarily fast WIMPs.

By employing Eqs. (\ref{eq:rate}) and (\ref{eq:ivmin}) one can derive
information on the quantity $\xi \sigma_{\rm scalar}^{(\rm nucleon)}$
from the measurements on the
differential rate $dR/dE_R$, once a specific velocity distribution function
$f(\vec{v})$ is selected.

\subsection{WIMP distribution functions}
\label{sec:df}

A systematic study of various distribution functions was carried  out in
Ref. \cite{bcfs} and a number of these DFs were subsequently
analyzed in Ref. \cite{constr} to study the sensitivity of the upper bounds on
$\xi \sigma_{\rm scalar}^{(\rm nucleon)}$
derived from different experiments of WIMP direct detection.

In Ref. \cite{bcfs}
 the phase--space DFs were classified into
four categories, depending on the symmetry properties of the matter
density (or the corresponding gravitational potential) and of the
velocity dependence: A) spherically symmetric matter density
$\rho_{\rm DM}$ with isotropic velocity dispersion, B) spherically
symmetric matter density with non--isotropic velocity dispersion, C)
axisymmetric models, D) triaxial models \cite{note0,others}.
For each category, different specific models are identified.

In Ref. \cite{bcfs} a procedure was developed to determine, for each individual
velocity DF, the relevant range of the $\rho_0$ values. Specifically,
each halo model was constrained by a number of observational inputs:
i) properties of the galactic rotational curve, namely
the range of the allowed values for the local rotational velocity,
$170~\mbox{km sec$^{-1}$}\leq v_0 \leq 270~\mbox{km sec$^{-1}$}$
\cite{kochanek,cepheids}, and the amount of flatness of the rotational
curve at large distances from the galactic center, and ii) the maximal
amount of non--halo components in the Galaxy, $M_{\rm vis}$ ({\em
i.e.} the disk, the bulge, etc.). These constraints determine the extremes
of the local dark matter density $\rho_0$. For instance, when one assumes
a maximal halo ({\it i.e.} the contribution of the
non-halo components is minimized) $\rho_0$ is maximal and,
on the contrary, the value of $\rho_0$ is minimal when  the contribution
of the halo to the rotational curve is minimal.
By this procedure, for any specific analytic form of the velocity DF and any value of
the local rotational velocity $v_0$ within the interval
$170~\mbox{km sec$^{-1}$}\leq v_0 \leq 270~\mbox{km sec$^{-1}$}$, one derives
the relevant lower and  upper bounds, $\rho_0^{\rm min}$ and $\rho_0^{\rm max}$, for the
local DM density. Obviously,   $\rho_0^{\rm min}$ and $\rho_0^{\rm max}$
are increasing functions of  $v_0$, since a large amount of local matter is necessary to
support a large value of the local rotational velocity.

Notice that both $\rho_0$ and $v_0$ are crucial parameters in establishing the properties
of the quantity $\xi \sigma_{\rm scalar}^{(\rm nucleon)}$, when this is extracted from the
detection rate. Whereas $\rho_0$ has the role of a normalization factor, $v_0$ is both related
to the WIMP kinetic energy and to the change in the reference frames, thus is crucial
in determining the shapes of the sensitivity levels usually reported in the plane WIMP
mass--$\xi \sigma_{\rm scalar}^{(\rm nucleon)}$.

In the present paper,
as our reference model we take the cored isothermal sphere,
which we will discuss in detail; we will also comment about some other DFs.
 The density profile of the
cored--isothermal sphere (denoted as Evans logarithmic model, or A1 model,  in
Ref.  \cite{bcfs}) is:

\begin{equation}
\rho(r) = \frac{v_0^2}{4 \pi G}\frac{3 R_c^2 + r^2}{(R_c^2 + r^2)^2}\, ,
\label{isot}
\end{equation}

\noindent
where $G$ is the Newton's constant, $v_0$ is the local value of the rotational velocity and
$R_c$ is the core radius. For $R_c$ we use the value $R_c = 5$ kpc.
For the parameter $v_0$ we will consider the values $v_0 = 170, 220, 270$ km sec$^{-1}$,
which represent the minimal, central and maximal values of $v_0$ in its physical range.
For these values of  $v_0$, the extreme values of $\rho_0$ are:
i)  $v_0 = 170$ km sec$^{-1}$ with
$\rho_0^{\rm min} = 0.20$ GeV cm$^{-3}$ and  $\rho_0^{\rm max} = 0.42$ GeV cm$^{-3}$;
ii) $v_0 = 220$ km sec$^{-1}$ with
$\rho_0^{\rm min} = 0.34$ GeV cm$^{-3}$ and  $\rho_0^{\rm max} = 0.71$ GeV cm$^{-3}$;
iii)  $v_0 = 270$ km sec$^{-1}$ with
$\rho_0^{\rm min} = 0.62$ GeV cm$^{-3}$ and  $\rho_0^{\rm max} = 1.07$ GeV cm$^{-3}$.

\subsection{Local fractional density}
\label{sec:resc}

The WIMP fractional density $\xi = \rho_W/\rho_0$ is
taken to be $\xi = {\rm min}\{1, \Omega_{\chi} h^2/(\Omega_{CDM}
  h^2)_{\rm min}\}$, in order to have rescaling \cite{gaisser},
when $\Omega_{\chi} h^2$ turns out to be less than $(\Omega_{CDM} h^2)_{\rm min}$ (here
$(\Omega_{CDM} h^2)_{\rm min}$ is set to the value 0.098,
as derived at a 2$\sigma$ level from
the results of Ref. \cite{wmap5}.)

\subsection{WIMP--nucleon cross section: hadronic uncertainties}
\label{sec:hadr}

The neutralino--nucleon scalar cross section $\sigma_{\rm scalar}^{(\rm nucleon)}$
is mainly due to exchanges of the two CP--even  neutral Higgs bosons, $h$ and $H$,
in the t--channel  \cite{barbieri} and to squark--exchanges in the s-- and u--channels \cite{griest}.
The expression of $\sigma_{\rm scalar}^{(\rm nucleon)}$ may be found in \cite{bdfs2}.

In the cross section representing the Higgs--exchanges, the factors involving the
couplings between the Higgs
bosons and the nucleon may be written as
$I_{h,H} = \sum_q k_q^{h,H} m_q \langle N|\bar{q} q |N\rangle$, where the coefficients
$k_q^{h,H}$ depend on supersymmetric parameters, $\langle N|\bar{q} q |N\rangle$
denotes the scalar density of the quark $q$ in the nucleon, and $m_q$ is the quark mass.
The coefficients $k_q^{h,H}$ are given in \cite{uncert2}.

The calculation of the quantities $ m_q \langle N|\bar{q} q |N \rangle $
is usually carried out by first expressing these quantities
in terms of the pion--nucleon sigma term
\be
\sigma_{\pi N} = \frac{1}{2} (m_u + m_d) <N|\bar uu + \bar dd|N>,
\ee
\noindent
of the quantity $\sigma_{0}$,
related to the size of the SU(3) symmetry breaking,
\be
\sigma_{0}\equiv \frac{1}{2}(m_u+m_d) <N|\bar uu+\bar dd-2\bar ss|N>,
\ee
\noindent and of the ratio $r=2 m_s/(m_u+m_d)$.

In fact, by assuming isospin invariance for quarks
$u$ and $d$,
the quantities $ m_q \langle N|\bar{q} q |N \rangle $ for light
quarks may be written as

\beqarr
m_u <N|\bar uu|N>&\simeq& m_d <N|\bar dd|N> \simeq \frac{1}{2} \sigma_{\pi N}
\label{eq:condlight}\\
m_s <N|\bar ss|N>&\simeq& \frac{1}{2} r (\sigma_{\pi N} - \sigma_{0}).
\label{eq:condstrange}
\eeqarr

As for the heavy quarks $c$, $b$, $t$, one conveniently
employs the heavy quark
expansion\cite{svz} to obtain

\beqarr
m_c <N|\bar cc|N> \simeq  m_b <N|\bar bb|N> \simeq m_t <N|\bar tt|N> \nonumber\\
\simeq \frac{2}{27} \left [ m_N - \sigma_{\pi N} +
\frac{1}{2}r (\sigma_{\pi N} - \sigma_{0})\right],
\label{eq:condheavy}
\eeqarr
\noindent where $m_N$ is the nucleon mass.

In this way the quantities
$I_{h,H}$ can be re--expressed as
\be
I_{h,H} = k_{u{\rm -type}}^{h,H} g_u + k_{d{\rm -type}}^{h,H} g_d,
\label{eq:i}
\ee
\noindent where
\beqarr
g_u &\simeq& m_l <N|\bar{l}l|N>+\;2\;  m_h <N|\bar{h}h|N> \nonumber\\
&\simeq& \frac {4} {27} (m_N + \frac {19}{8} \sigma_{\pi N}
- \frac{1}{2}r (\sigma_{\pi N} - \sigma_{0})),
\label{eq:l}
\eeqarr
\beqarr
  g_d &\simeq& m_l<N|\bar{l}l|N>+\; m_s<N|\bar{s}s|N>+  \nonumber \\
  & & +\; m_h<N|\bar{h}h|N> \nonumber\\
&\simeq& \frac{2}{27} (m_N + \frac{23}{4} \sigma_{\pi N}
+ \frac{25}{4} r  (\sigma_{\pi N} - \sigma_{0})); \label{eq:g}
\eeqarr

\noindent
 $l$ stands for light quarks ($l=u,d$) and $h$ denotes the heavy
 ones ($h=c,b,t$).

The major problem here is that the three quantities $\sigma_{\pi N}$,
$\sigma_{0}$ and $r$ are all affected by sizable uncertainties which
in turn dramatically affect the determination of the coefficients
$g_u$ and $g_d$. This problem was stressed in Ref.  \cite{uncert1,uncert2} (see also
Refs. \cite{noimpl,jkg} for  earlier discussions of this point) and subsequently
also recognized by other authors \cite{ellis,arnowitt,nath,fmw,eos}.

The range of $\sigma_{\pi N}$ we considered in Ref. \cite{uncert1}
was

\be
41\; {\rm MeV} \, \lsim \sigma_{\pi N}\lsim 57 \; {\rm MeV}
\label{eq:k}
\ee

\noindent
 as derived from the
pion--nucleon scattering amplitude, calculated at the so--called
Cheng--Dashen point by Koch \cite{koch1}, and from the evolution of the
nucleon scalar form factor,  as a function of the momentum
transfer from $t=2 m_{\pi}^2$ to $t=0$, evaluated in Ref. \cite{gls2}.

In Ref. \cite{uncert2} we reconsidered the calculation of the
coefficients $g_u, g_d$ in  light of a new determination
of $\sigma_{\pi N}$ presented in
Ref. \cite{pavan}. In fact, the George Washington University/TRIUMF group,
using   up--dated pion--nucleon scattering data
\cite{said} and  a new partial--wave and dispersion relation analysis
program,  derived a  range for $\sigma_{\pi N}$ \cite{pavan}

\be
55 \; {\rm MeV} \lsim \sigma_{\pi N}\lsim 73 \; {\rm MeV},
\label{eq:p}
\ee

\noindent
which turned out to be sizeably larger than the one of
Eq. (\ref{eq:k}).
Values of the nucleon scalar form factor at the Cheng--Dashen point
higher than those of Ref. \cite{koch1} were also reported in
Ref. \cite{olsson}.

     In Ref. \cite{uncert2} we stressed the dramatic importance that
the uncertainties in the hadronic quantities $\sigma_{\pi N}$,
$\sigma_0$ and $r$ have in a number of fundamental issues, namely:
i) evaluation of WIMP detection rates in direct and also in some
indirect searches (neutrino fluxes due to WIMP self--annihilation in
Earth and Sun), ii) actual regions of the supersymmetric parameter space
involved in searches for WIMPs, iii) connection between experimental event
rates and relic abundance.

     It is unfortunate that the dichotomy between the two determinations
in Eq. (\ref{eq:k}) and in Eq. (\ref{eq:p}) still persists. This fact is not only
related to the experimental determination of the relevant quantities in the pion--nucleon
scattering but also to the intricacies involved in the derivation of
$\sigma_{\pi N}$ from the experimental data. It is worth noting that the
uncertainty inherent in the value of $\sigma_{\pi N}$ could be even larger than the
one exemplified by Eqs. (\ref{eq:k})--(\ref{eq:p}) (see Ref. \cite{clg} for further
determinations).

    For definiteness, in the present paper we consider variations of $\sigma_{\pi N}$
in the range which is the union of the two intervals of
Eqs. (\ref{eq:k})--(\ref{eq:p}), {\it i.e.}

\be
41 \; {\rm MeV} \lsim \sigma_{\pi N}\lsim 73 \; {\rm MeV}.
\label{eq:q}
\ee

In the presentation of our results, we will also consider a
reference point, representative of a value of
$\sigma_{\pi N}$ which is within the narrow overlap of the two
ranges of Eqs. (\ref{eq:k})--(\ref{eq:p}).

The quantity ${\sigma_{0}}$ is taken in the range \cite{gl}

\beq
\sigma_{0}=30\div40 \;{\rm MeV}
\label{eq:0}
\eeq

\noindent
and to the mass ratio $r= 2 m_s/(m_u+m_d)$
the {\it default} value $r = 25$ is assigned.

We recall that
the fractional strange--quark content of the nucleon $y$
\be \label{eq:y}
y=2 \frac{<N|\bar ss|N>}
{<N|\bar uu+ \bar dd|N>},
\ee

\noindent
is linked to $\sigma_{0}$ and $\sigma_{\pi N}$ by the expression

\be
y=1-\frac{\sigma_{0}}{\sigma_{\pi N}}.
\label{eq:yy}
\ee

The reference point mentioned above, meant to represent an estimate of the
hadronic quantities in agreement with both ranges of Eq.(\ref{eq:k}) and
Eq.(\ref{eq:p}),
is  defined by the values $g_{u,ref}$ = 123 MeV, $g_{d,ref}$ = 290 MeV
(this set of values is the one  employed also in our
previous paper of Ref. \cite{zooming}). As mentioned above, together with
this representative point, we will also explicitly discuss the
implications of the full range of $\sigma_{\pi N}$ given in Eq.(\ref{eq:q}).

\section{The annual modulation effect measured by the DAMA Collaboration}
\label{sec:libra}

The first results of the DAMA/LIBRA experiment on direct detection
of dark matter  particles have recently been presented
\cite{dama/libra}. These data concern an exposure of 0.53 ton yr.
When added to the previous exposure of 0.29 ton yr of the DAMA/NaI
experiment \cite{dama/nai},  the total exposure collected by the
DAMA Collaboration, with the DAMA/LIBRA and the DAMA/NaI
experiments together, amounts to 0.82 ton yr. The analysis of the
total set of data shows an annual modulation effect in the event
rate at 8.2 $\sigma$ C.L. This yearly modulation satisfies all
the features expected for an annual variation due to relic
particles in our galactic halo \cite{freese} and is not
explained by systematic effects and/or seasonal variations of
other various origins \cite{dama/libra}.

\begin{figure}[t] \centering
\vspace{-20pt}
\includegraphics[width=1.0\columnwidth]{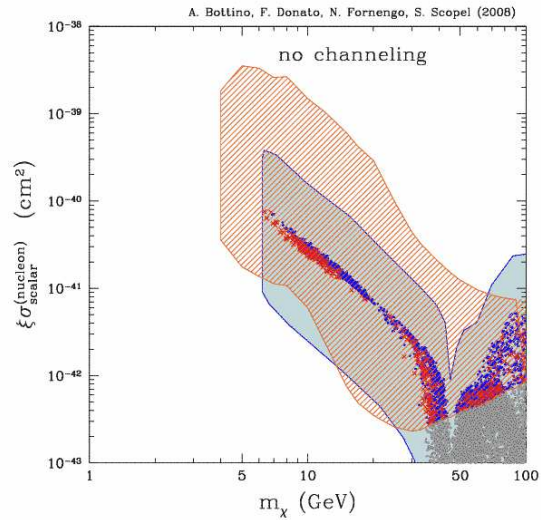}
\vspace{-30pt}
  \caption{$\xi \sigma_{\rm scalar}^{(\rm nucleon)}$ as a function of the WIMP mass.
The region covered by a (red) slant hatching denotes the DAMA annual modulation region,
under the hypothesis that the effect is due to a WIMP with a coherent interaction with nuclei
and {\it without including} the channeling effect. This region represents the domain where the
likelihood-function values
differ more than 6.5 $\sigma$ from the null hypothesis (absence of modulation).
It has been derived by the DAMA Collaboration by varying the WIMP
galactic distribution function  over the set considered in Ref.\cite{bcfs} and
by taking into account other uncertainties of different origins \cite{note1}.
 The scatter plot
  represents supersymmetric configurations calculated with the  model
  summarized in  Sect. \ref{sec:susy}, at the fixed
representative set of values for the hadronic quantities characterized by:
$g_{u,ref}$ = 123 MeV, $g_{d,ref}$ = 290 MeV.
  The (red) crosses denote configurations
  with a neutralino relic abundance which matches the WMAP cold
  dark matter amount ($0.098 \leq \Omega_{\chi} h^2 \leq 0.122$), while the
  (blue) dots refer to configurations where the neutralino is subdominant
  ($\Omega_{\chi} h^2 < 0.098$). The (blue) uniformly--shaded region  represents
the extension of the scatter plot upwards and downwards, when the hadronic uncertainties
reported in Eq. (\ref{eq:q}) are included (see text).}
\label{fig:01}
\end{figure}

\begin{figure}[t] \centering
\vspace{-20pt}
\includegraphics[width=1.0\columnwidth]{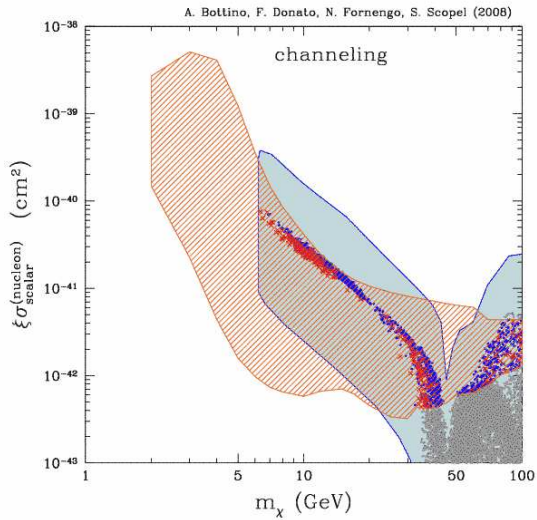}
\vspace{-30pt}
  \caption{$\xi \sigma_{\rm scalar}^{(\rm nucleon)}$ as a function of the WIMP mass.
The region covered by a (red) slant hatching denotes the DAMA annual modulation region,
under the hypothesis that the effect is due to a WIMP with a coherent interaction with nuclei
and {\it including} the channeling effect. All other prerequisites of this region are as in
Fig. \ref{fig:01}.
 The scatter plot and  the (blue) uniformly--shaded region are as in Fig. \ref{fig:01}.}
\label{fig:02}
\end{figure}

The relic particles responsible for the DAMA annual modulation
effect can be of many different kinds; a number of possible
candidates are recollected in Ref. \cite{dama/libra} (see {\it e.g.} 
also \cite{dama_altre,weiner,foot,sneutrino}). In the
present paper we analyze the interpretation of the DAMA effect in
terms of relic neutralinos.

Indeed, we already invoked neutralinos  \cite{noi97} since the very first DAMA
results of Ref. \cite{dama97} which showed an yearly variation in
the direct signal. Our interpretation was further
pursued, most notably in Refs. \cite{lowneu,lowdir,ind}, where the
focus was set on light neutralinos ({\it i.e.} neutralino with masses
$m_{\chi} \lsim$ 50 GeV).

More recently, we presented a new investigation of the DAMA/NaI results \cite{zooming},
when the DAMA Collaboration discussed the possible implications of including
the channeling effect in the analysis of their data \cite{channeling}.

In this paper we update our interpretation of the annual
modulation effect in  light of the experimental results
improved by the quite sizable increase in total exposure of the recent
DAMA experiments.

From now on, when mentioning DAMA data we will mean the total set of data including
both the DAMA/NaI experiment and the DAMA/LIBRA one, reported in Refs. \cite{dama/libra}.

In the following section we will discuss
the annual modulation regions, in the plane
$m_{\chi}$--$\xi \sigma_{\rm scalar}^{(\rm nucleon)}$ ($m_{\chi}$ denotes the WIMP mass),
that have been derived by the DAMA Collaboration for the case of
a WIMP with a coherent interaction with nuclei \cite{dama3}.
The derivation has been carried out both in the case where the channeling effect is taken
into account and in the one where channeling is not included.
 We recall that the features of the annual modulation region in the plane
$m_{\chi}$--$\xi \sigma_{\rm scalar}^{(\rm nucleon)}$ depend sensitively on
 the specific model--dependent
procedure employed in the evaluation of the channeling effect.
The regions reported hereby for the case of channeling
 correspond to the case where channeling
is included with the model explained in Ref. \cite{channeling}.
Actually, the extent by which the channeling effect occurs when a putative WIMP
traverses a NaI crystal is still under study. For this reason, in our analysis we
consider both cases of no-channeling and of channeling with the model of Ref. \cite{channeling}.
One expects that the actual physical situation is comprised within
these two cases.

\subsection{Annual modulation regions (convolution over a class of distribution functions)}
\label{sec:regions_con}

We start our analysis by showing in Figs. \ref{fig:01}--\ref{fig:02} the DAMA annual modulation regions in a plot
of $\xi \sigma_{\rm scalar}^{(\rm nucleon)}$ versus the WIMP mass.
   These have been derived by the DAMA Collaboration for the case of
   a WIMP with a coherent interaction with nuclei \cite{dama3}, by varying the WIMP
galactic distribution function (DF) over the set considered in Ref.\cite{bcfs} and
by taking into account
other uncertainties of different origins \cite{note1}.
The DAMA regions
are denoted by a (red) slant hatching; they
represent regions where the likelihood-function values differ more than
6.5 $\sigma$ from the null hypothesis (absence of modulation).
Fig. \ref{fig:01} refers to the case
in which the channeling effect  is not included, whereas Fig. \ref{fig:02}
displays the case where channeling is included.

Figs. \ref{fig:01}--\ref{fig:02} also show the (blue)
uniformly-shadowed region, which represents the physical neutralino region as derived within
our effective MSSM. The scatter plot, common to Fig.1 and Fig.2, denotes the results of our evaluations,
when a scanning over the supersymmetric parameter space is performed, at the fixed
representative set of values for the hadronic quantities
mentioned in Sect. \ref{sec:hadr}. This representative set is characterized by the
values: $g_{u,ref}$ = 123 MeV, $g_{d,ref}$ = 290 MeV (we recall that this set is the one  employed also in our
previous paper of Ref. \cite{zooming}). The (red) crosses and the
(blue) dots of the scatter plot denote configurations with no-rescaling and those
with rescaling of the local density, respectively (see Sect. \ref{sec:resc}).
The features of the supersymmetric configurations belonging to the 
scatter plot are discussed in detailed in Ref. \cite{lowneu}. 

The uniformly--shaded region displayed in Figs. \ref{fig:01}--\ref{fig:02} represents
the extension of the scatter plot upwards and downwards, when the hadronic uncertainties
reported in Eq. (\ref{eq:q}) are included. The range of the pion--nucleon sigma term of
Eq. (\ref{eq:p}) is responsible for the upper extension of the physical region, as compared
to the representative scatter plot, by an enhancement factor of about 2--3, whereas
the range of Eq. (\ref{eq:k}) generates the lower extension by a suppression factor of order 8--9.
These numbers follow immediately from the formulae in Sect. \ref{sec:hadr}, by taking into account that
the dominant term in the quantity $I_{h,H}$ is the one involving $g_d$.
Thus, the scatter plot for any value set of hadronic quantities with  given values
of $g_d$ and $g_u$
can  {\it approximately} be obtained from the one corresponding to the reference set of values,
characterized by $g_{u,ref}$ = 123 MeV, $g_{d,ref}$ = 290 MeV,
by scaling the reference scatter plot by the factor $(g_d/g_{d,ref})^2$. However, notice that, in
deriving the boundaries of the full theoretical region in Figs. \ref{fig:01}--\ref{fig:02},
the full expression of Eq. (\ref{eq:i}) has been used.

From Figs.1--2 it is clear that the DAMA annual modulation region is largely compatible
with the theoretical predictions for relic neutralinos with masses
$m_{\chi} \lsim$ 100 GeV, in particular for neutralinos within the low--energy funnel
for $m_{\chi} \lsim$ 50 GeV. This occurs, whether or not the channeling effect is included.

\subsection{Annual modulation regions for single  halo models}
\label{sec:regions_mod}

We turn now to the analysis of the annual modulation regions for specific forms of the
WIMP distribution function. First we discuss in detail our reference model,
the cored isothermal sphere, mentioned in Sect. \ref{sec:df}
(denoted as Evans logarithmic model, or A1 model,  in
Ref.  \cite{bcfs});
we will comment about some other DFs afterwords.

\begin{figure*}[t] \centering
\vspace{-20pt}
\includegraphics[width=2.0\columnwidth]{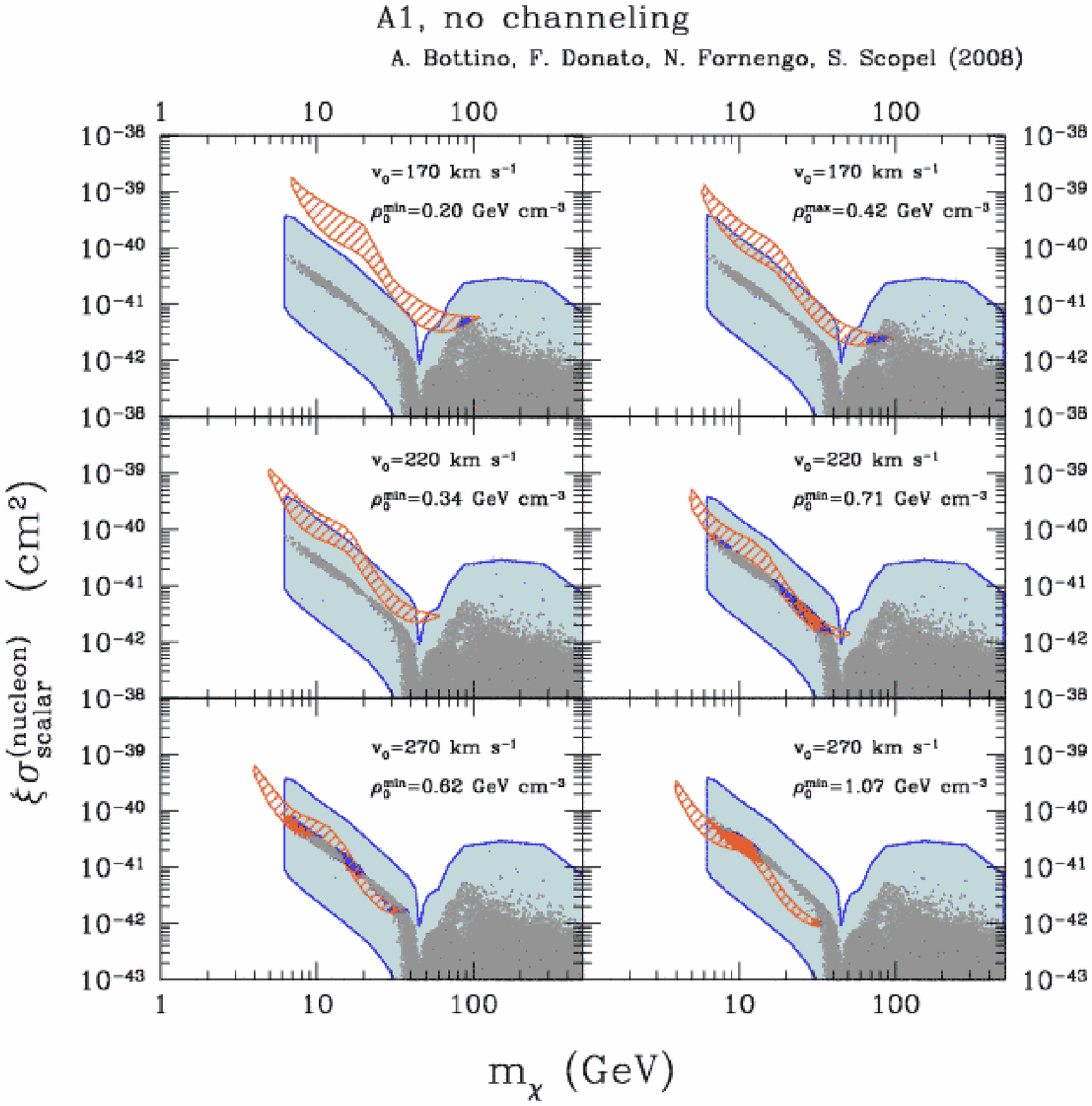}
\vspace{-30pt}
  \caption{$\xi \sigma_{\rm scalar}^{(\rm nucleon)}$ as a function of the WIMP mass.
The region covered by a (red) slant hatching denotes the DAMA annual modulation region,
under the hypothesis that the effect is due to a WIMP with a coherent interaction with nuclei
and {\it without including} the channeling effect. This region represents the domain where the
likelihood-function values
differ more than 6.5 $\sigma$ from the null hypothesis (absence of modulation).
It has been derived by the DAMA Collaboration by assuming that the WIMP distribution function is
given by the cored isothermal sphere  (denoted as Evans logarithmic model, or A1 model,  in
Ref.  \cite{bcfs}) and using the parameters of set A of Sect. 7.2 of the first paper of
Ref. \cite{dama/nai}. The scatter plot and  the (blue) uniformly--shaded region are as
in Fig. \ref{fig:01}.}
\label{fig:03}
\end{figure*}

\begin{figure*}[t] \centering
\vspace{-20pt}
\includegraphics[width=2.0\columnwidth]{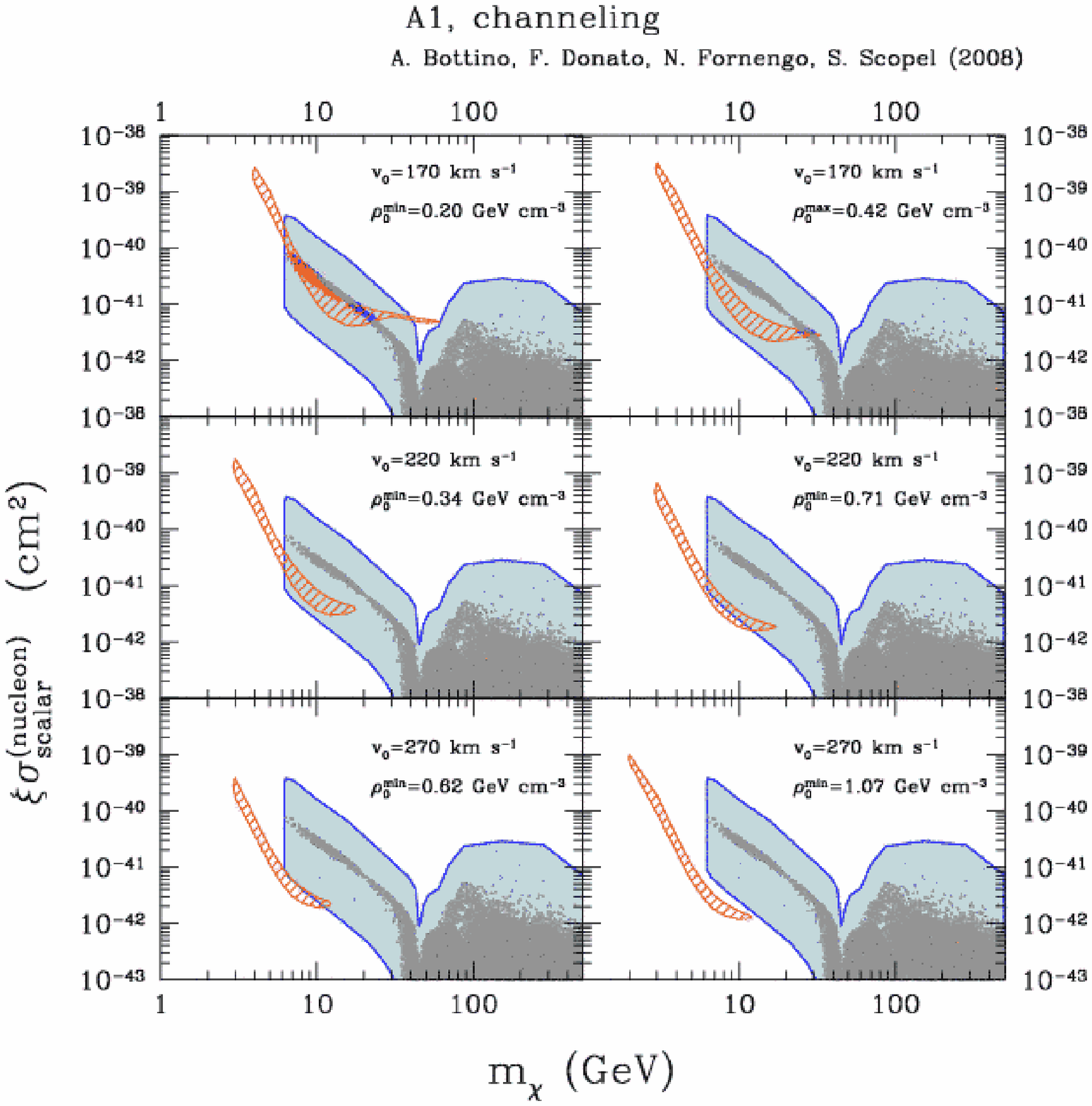}
\vspace{-30pt}
  \caption{$\xi \sigma_{\rm scalar}^{(\rm nucleon)}$ as a function of the WIMP mass.
The region covered by a (red) slant hatching denotes the DAMA annual modulation region,
under the hypothesis that the effect is due to a WIMP with a coherent interaction with nuclei
and {\it  including} the channeling effect.
All other prerequisites of this region are as in
Fig. \ref{fig:03}. The scatter plot and  the (blue) uniformly--shaded region are as
in Fig. \ref{fig:01}.}
\label{fig:04}
\end{figure*}

 Figs.\ref{fig:03}--\ref{fig:04} display  the theoretical predictions of our supersymmetric model
 (already shown in Figs.\ref{fig:01}--\ref{fig:02})
 together with the DAMA annual modulation regions \cite{dama3},
  under the hypothesis that the WIMP-nucleus interaction is coherent and that the
  velocity DF is given by the  cored isothermal sphere. The various insets refer to the
  representative values of $v_0$ and $\rho_0$ discussed before, that is:
i)  $v_0 = 170$ km sec$^{-1}$ with
$\rho_0^{\rm min} = 0.20$ GeV cm$^{-3}$ and  $\rho_0^{\rm max} = 0.42$ GeV cm$^{-3}$;
ii) $v_0 = 220$ km sec$^{-1}$ with
$\rho_0^{\rm min} = 0.34$ GeV cm$^{-3}$ and  $\rho_0^{\rm max} = 0.71$ GeV cm$^{-3}$;
iii)  $v_0 = 270$ km sec$^{-1}$ with
$\rho_0^{\rm min} = 0.62$ GeV cm$^{-3}$ and  $\rho_0^{\rm max} = 1.07$ GeV cm$^{-3}$.
 $v_{esc}$ is set to the value $v_{esc}$ = 650 km sec$^{-1}$.

 These DAMA annual modulation regions
represent domains  where the likelihood-function values differ more than
6.5 $\sigma$ from the null hypothesis (absence of modulation).
 The notations for the various regions and for the scatter plot are the same as in
 Figs. \ref{fig:01}--\ref{fig:02};
 Fig. \ref{fig:03} refers
to the case in which the channeling effect is not included, and Fig. \ref{fig:04}
to the case where
channeling is included.

It is remarkable that relic neutralinos are able to provide a good fit to the
experimental data.
In case of no-channeling,  low values of $v_0$ and $\rho_0$
($v_0 \simeq$ 170 km sec$^{-1}$ and $\rho_0 \simeq 0.2$ GeV cm$^{-3}$)
appear to be somewhat disfavored, though in this case neutralinos with
$m_{\chi} \simeq$ 60--100 GeV could be involved.  The agreement between
experimental data and theoretical evaluations in our model looks very good
for the other combinations of $v_0$ and $\rho_0$ values,
 with an
overall preference for neutralinos of low mass.
For the case where the channeling is included according to the modelling of
Ref. \cite{channeling}, experimental data favor values
of  $v_0$ and $\rho_0$ which are in the low-medium
side of their own physical ranges, {\it i.e.} $v_0 \simeq$ (170 -- 220) km sec$^{-1}$
and $\rho_0 \simeq (0.3 - 0.4)$ GeV cm$^{-3}$ and neutralino masses
in the mass range $m_{\chi} \simeq (7 - 30)$ GeV.

When no rotation of the halo is
considered,  the features of the annual modulation
region in the $m_{\chi} -  \xi \sigma^{\rm nucleon}_{\rm scalar}$ plane do not
differ much  when the galactic DF is varied among many of the galactic DFs
considered in Ref. \cite{bcfs}. For instance, for a matter density with a
Navarro-Frenk-White profile (A5 model of Ref. \cite{bcfs}) or for an isothermal
model with a non-isotropic velocity dispersion (B1 model of Ref. \cite{bcfs})
the physical situations are very similar to the ones depicted in Fig.3--4.

However, in the case  of DFs with triaxial spatial distributions (within the class D  of
Ref. \cite{bcfs}) and for models with a co-rotating halo there can be an
elongation of the annual modulation region towards heavier masses (these are
generic characteristic
features which can be derived from the analysis of Ref. \cite{bcfs}).
These features are displayed in Figs. 1-2 for masses up to the value of 100 GeV, 
which corresponds to the upper extreme of the annual--modulation 
region provided up to now by the DAMA Collaboration.

We wish to recall that the distribution
of WIMPs in the Galaxy could deviate from the models mentioned above, mainly
because of the presence of streams. For modification of the annual modulation
region in these instances see Ref.\cite{dama/nai,sat}.

\section{Upper bounds on the WIMP-nucleon scalar cross section}
\label{sec:bounds}

 Other experiments of WIMP direct detection,
 different from DAMA/NaI and DAMA/LIBRA, do not currently have the capability of measuring
 the annual modulation effect; they can only provide upper bounds for the expected
 signals  \cite{taup2007,xe,cdms,kims}. Once a specific form for the WIMP DF is
 taken and the relevant parameters fixed,
 these bounds can be converted into constraints on the WIMP-nucleus cross section.
 It is obvious that, in order to derive solid constraints on the WIMP physical properties,
 a conservative approach has to be followed in the selection of the WIMP distribution function;
 {\it i.e.} among the wide variety of DFs, one has to select the ones which imply the
 smallest responses in the detector.

\begin{figure} \centering
\vspace{-40pt}
\includegraphics[width=1.1\columnwidth]{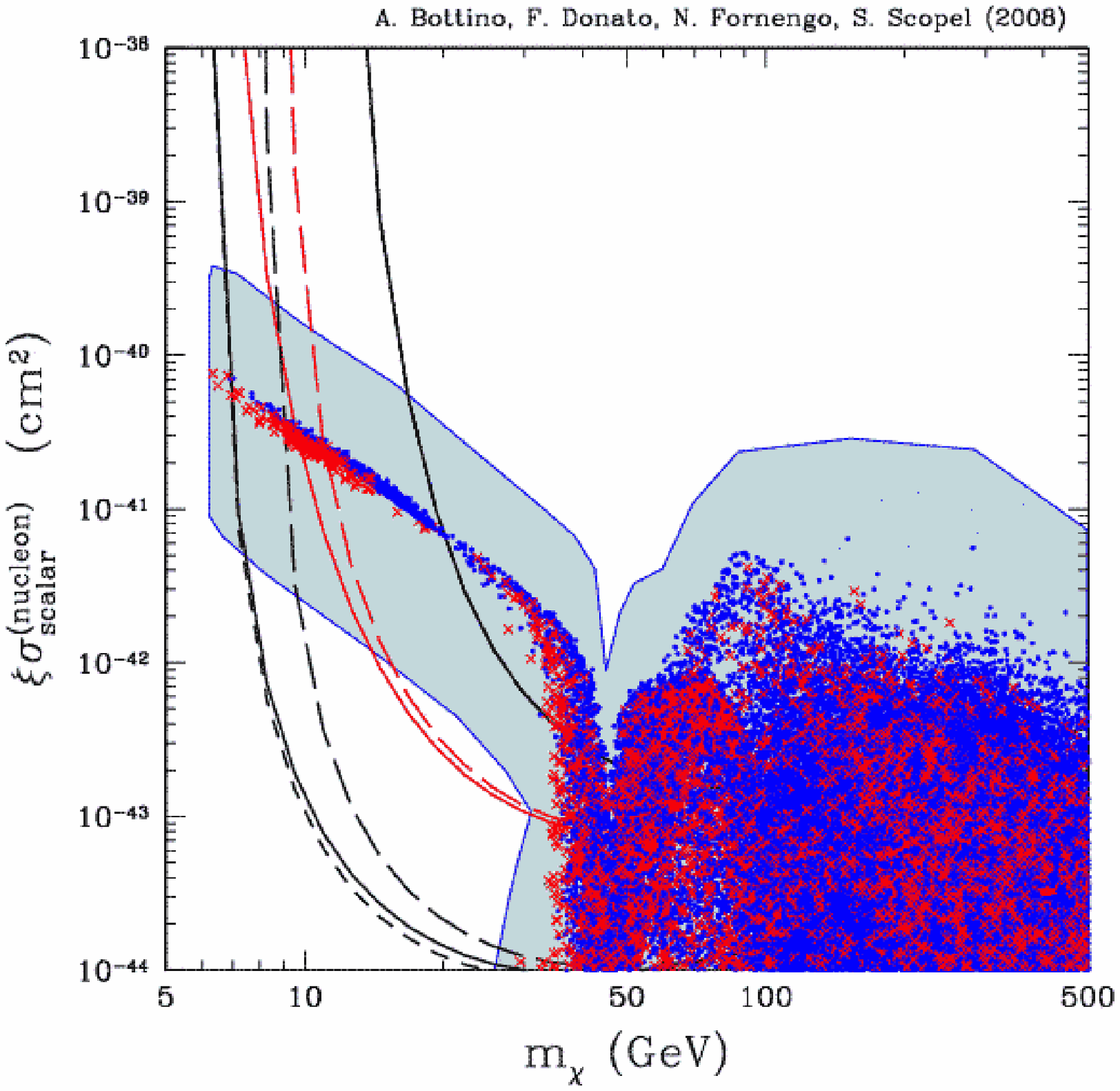}
\vspace{-30pt}\caption{The solid lines show
 the upper limit on the quantity $\xi \sigma_{\rm scalar}^{(\rm nucleon)}$
 as a function of the WIMP mass $m_\chi$ for the CDMS detector \cite{cdms}
and for $\vesc=650$ km sec$^{-1}$. The (red) median line refers to the
standard isothermal sphere with $v_0 =220$ km sec$^{-1}$ and $\rho_0
=0.3 ~{\rm GeV\,cm^{-3}}$ (model A0 of Ref. \cite{bcfs}). The (black) upper and lower curves
 refer to model B1 with
$v_0 =170$ km sec$^{-1}$ (upper solid line) and model C3 with $v_0
=270$ km sec$^{-1}$ (lower solid line). The short--dashed line refers to
model C3 with maximal counter--rotation of the galactic halo.
 The long--dashed lines show the upper limits for CDMS in the
case of a lower escape velocity $\vesc=450$ km sec$^{-1}$: the upper
line refers to model A1, the lower one to model C3. For model B1, the
limit coincides with the corresponding solid line.
The scatter plot and  the (blue) uniformly--shaded region are as
in Fig. \ref{fig:01}. Other specifications in the text.}
\label{fig:05}
\end{figure}

  As an example, we take the data recently published by the CDMS Collaboration \cite{cdms}.
 From these data, applied to a WIMP with a coherent interaction with nuclei,
 one can  derive the upper bounds for the quantity $\xi \sigma_{\rm scalar}^{(\rm nucleon)}$
 displayed in Fig. \ref{fig:05}, for a number of various DFs. The strong dependence of the
 upper bounds on the assumed DF is apparent. Superimposed to these limits
 is our theoretical region for relic neutralinos. We note that the conservative upper bound,
established by the B1 distribution function, does not exclude any
of the light neutralino configurations, when the hadronic
uncertainties are taken into account (only some configurations with neutralino
masses above 60 GeV are excluded). Of course,
other WIMP distribution functions, such as the isothermal sphere,
would introduce constraints on the supersymmetric population.

 However, one has to notice that the aforementioned bounds are obtained through involved
 procedures for discriminating electromagnetic signals from recoil events and through delicate
 subtractions meant to separate putative WIMP signals from neutron-induced events. A major
 critical point in these experiments and related analyses is that the very signature
 (the annual modulation) of the searched signal cannot be employed in extracting the authentic events.
Possible problems of stability in the acquisition parameters can affect the rejection procedures applied
to a large number of events, as well as the determination of the threshold and of the
energy scale \cite{dama/noblegas}.
In view of possible sizable uncertainties involved in these
procedures,
 we conservatively do not implement the
upper bounds discussed in this  section, while comparing theoretical expectations for relic
neutralinos to the annual modulation data of Ref. \cite{dama/libra}.

 It is also worth noting  that the upper bounds of Ref.
 \cite{cdms}, even when taken at their face value, are however not in conflict with the annual
 modulation data {\it and} the theoretical neutralino interpretation for
 masses $\sim  7-10$ GeV.
Compatibility of the DAMA/LIBRA annual--modulation results 
with the upper bounds of other experiments for coherently--interacting WIMPs 
with the masses $\lsim$ 10 GeV is also found in two papers \cite{feng, zurek},
appeared concomitantly with the present article. 

Finally, we recall that another experiment (KIMS \cite{kims}), running at present with a
detector
of about 104 kg of CsI crystals, is meant to provide a measurement of
the WIMP annual modulation in the future.

\begin{figure*}[t] \centering
\vspace{-20pt}
\includegraphics[width=2.0\columnwidth]{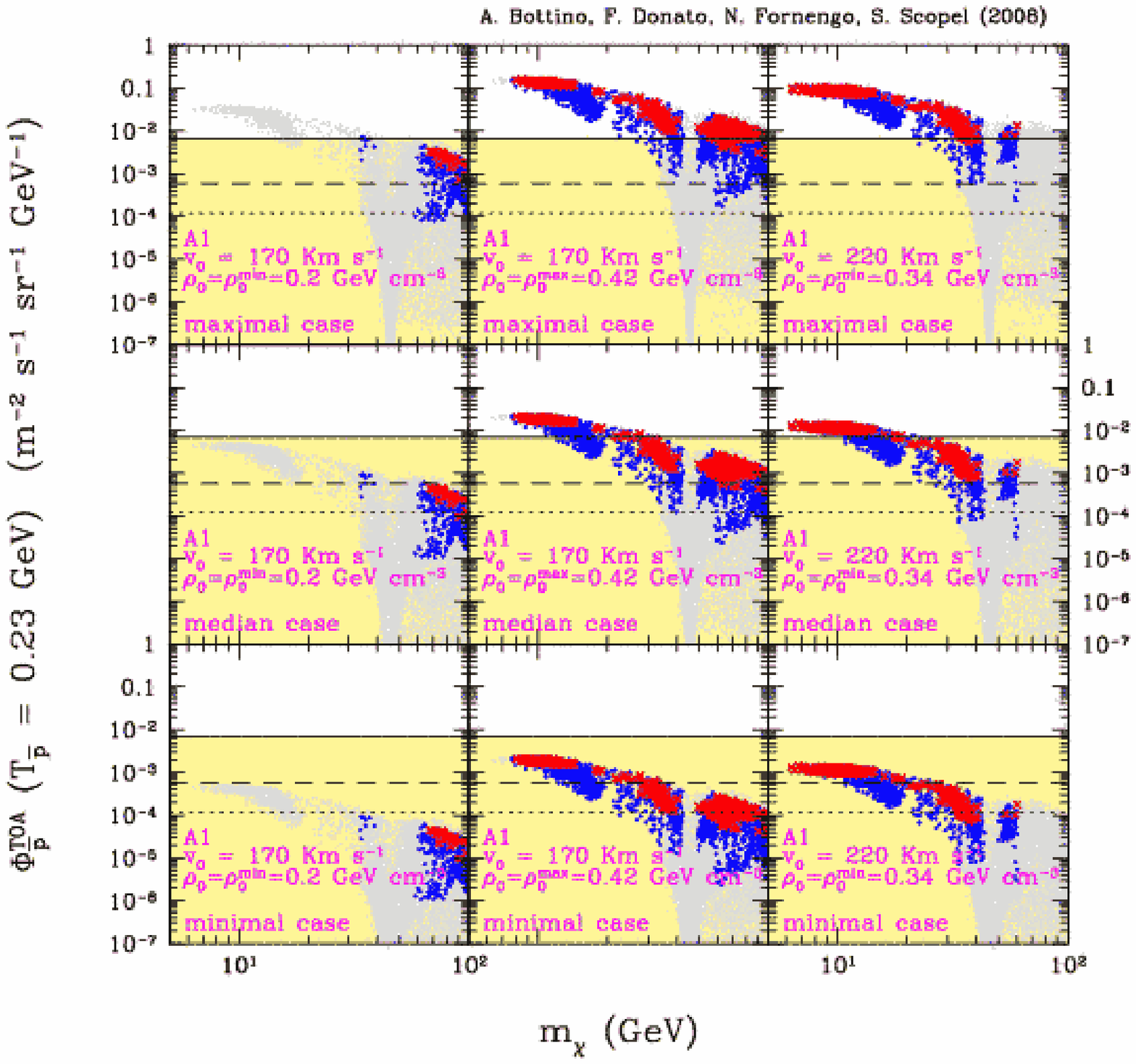}
\vspace{-30pt}
  \caption{Antiproton flux, at $\bar p$ kinetic energy $T_{\bar p}=0.23$ GeV,
  generated by the neutralino
  configurations selected by the DAMA data, when
   a cored isothermal halo is employed and
  the channeling effect is not included.
  Each row
  corresponds to a different set of cosmic rays propagation parameters: the
  upper, median and lower rows refer to the sets which provide the maximal,
  median and minimal antiproton flux, respectively (see Table I).
The three columns refer to the sets of $v_0$ and $\rho_0$ values denoted as sets $A$, $B$ and
$C$ in the text.
  The bold (colored) points refer to configurations
  compatible with the DAMA regions  for the case in which the channeling effect
  is not included (see Fig. \ref{fig:03}). In selecting the allowed configurations the hadronic
  uncertainties have been taken into account. The points are
  differentiated as follows: (red) crosses denote configurations with a
  neutralino relic abundance which matches the WMAP cold dark matter amount
  ($0.098 \leq \Omega_{\chi} h^2 \leq 0.122$), while (blue) dots refer to
  configurations where the neutralino is subdominant ($\Omega_{\chi} h^2 <
  0.098$). The light gray points denote configurations with a
  neutralino--nucleon scattering cross section outside the corresponding
  DAMA allowed region.
  The solid horizontal line shows the maximal allowable amount of
  antiprotons in the BESS data \cite{bess} over the secondary component; the
  dashed and dotted lines denote estimates of the PAMELA \cite{pamela} and
  AMS-02 \cite{ams} sensitivities
  to exotic antiprotons for 3 years missions, respectively.}
\label{fig:06}
\end{figure*}

\begin{figure*}[t] \centering
\vspace{-20pt}
\includegraphics[width=2.0\columnwidth]{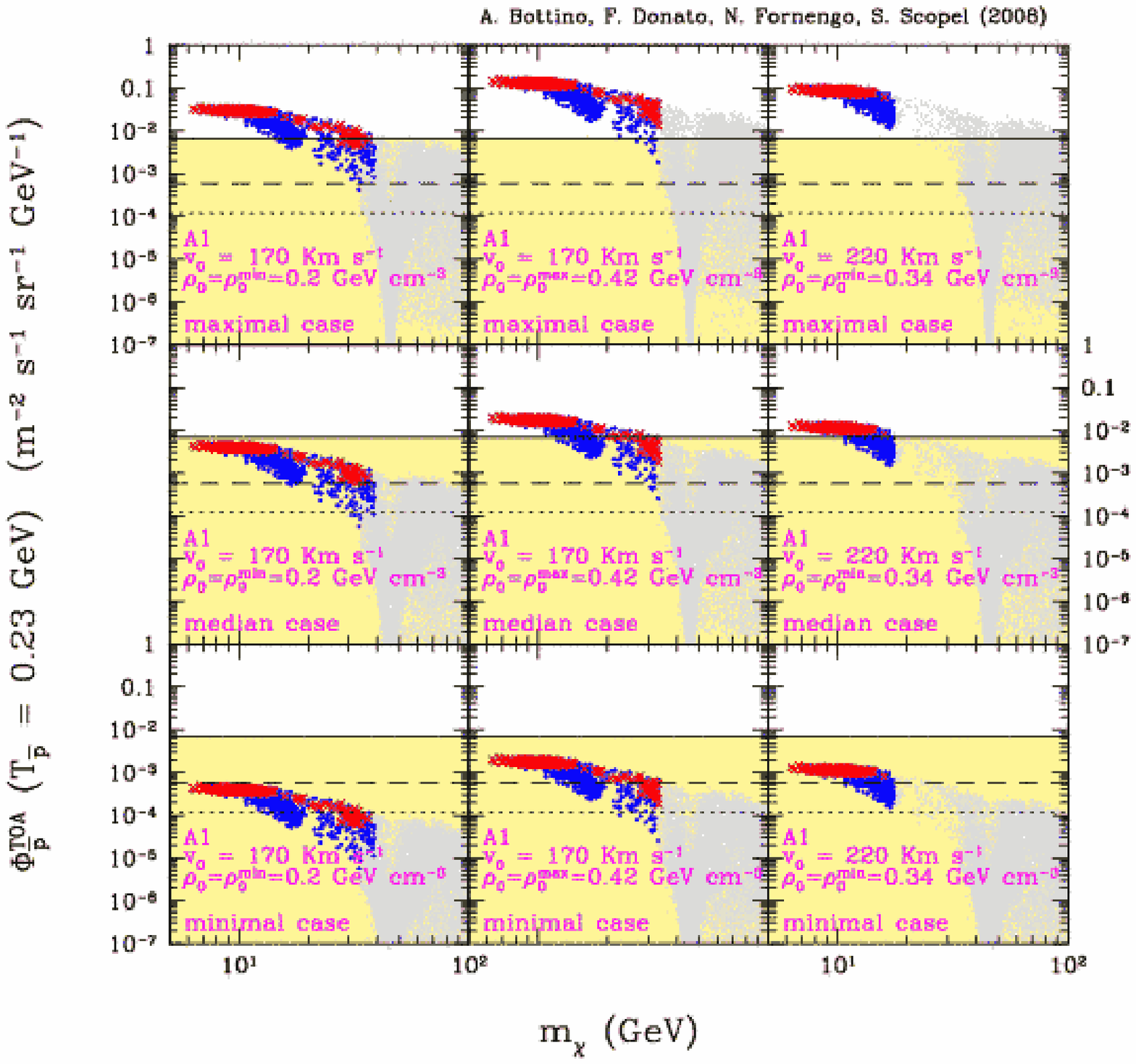}
\vspace{-30pt}
  \caption{Antiproton flux, at $\bar p$ kinetic energy $T_{\bar p}=0.23$ GeV,
   generated by the neutralino
  configurations selected by the DAMA data, when
   a cored isothermal halo is employed and
  the channeling effect is included.
   Notations are as in Fig. \ref{fig:06}, except that here
  the neutralino configurations of the scatter plot are those selected by the
  DAMA regions of  Fig. \ref{fig:04}.}
\label{fig:07}
\end{figure*}

\section{Galactic antiprotons}
\label{sec:antiprotons}

In Sects. \ref{sec:regions_con}--\ref{sec:regions_mod}
 we have seen that the agreement of the DAMA data with theoretical
evaluations for relic neutralinos is quite good for light neutralino masses and
for wide intervals of the astrophysical
quantities $\rho_0$ and $v_0$ within the ranges discussed in Sect. \ref{sec:df}.
However, it turns out that  the data of
cosmic antiprotons fluxes can put stringent limits on light neutralino configurations
\cite{bdfs,zooming}.

   In fact, in Ref. \cite{don} it is shown that the experimental antiproton spectrum is fitted
very well by the secondary component from cosmic rays spallation,
calculated with the set of the diffusion parameters which is derived from the
analysis of the boron--to--carbon ratio (B/C) component of cosmic rays \cite{PaperI}. Indeed the
calculated secondary flux of cosmic antiprotons fits the experimental data with
with a ${\chi}^2$ = 33.6 with 32 data points, and with an uncertainty of 20\%.
 This means that very little room is left for  a possible primary component of antiprotons
 generated by an exotic origin (neutralino self--annihilation in our case).

Let us recall that in Ref. \cite{don} the
secondary antiproton spectrum,
generated by spallation processes, was propagated using a two--zone diffusion model described in terms
of five parameters. Two of these parameters, $K_0$ and $\delta$, enter the expression of the
diffusion coefficient: $K = K_0 \beta R^{\delta}$ ($R$ is the particle rigidity);
the other three parameters are the Alf\'en
velocity $V_A$, the velocity of the convective wind $V_c$, and  the thickness
$L$ of the two large diffusion layers which sandwich the thin galactic disk \cite{PaperI}.
When studying the primary antiprotons one can extract the following  three
sets of propagation parameters: the best--fit (on B/C) set (denoted as median),
together with the sets which yield the minimal and the maximal primary antiproton fluxes
\cite{pbar_susy}.
The values of these three sets  are given in Table I.

\begin{table}[t]
\begin{center}
{\begin{tabular}{@{}|c|c|c|c|c|c|@{}}
\hline
~~{\rm case}~~ &  ~~~~$\delta$~~~~  & $K_0$                 & $L$   & $V_{c}$       &
$V_{A}$
\\  & & ~[${\rm kpc^{2}/Myr}$]~ & ~[kpc]~ & ~[km s$^{-1}$]~ & ~[km s$^{-1}$]~ \\
\hline
\hline
{\rm max} &  0.46  & 0.0765 & 15 & 5    & 117.6  \\
{\rm med} &  0.70  & 0.0112 & 4  & 12   &  52.9  \\
{\rm min} &  0.85  & 0.0016 & 1  & 13.5 &  22.4  \\
\hline
\end{tabular}}
\caption{ Astrophysical parameters of the two--zone diffusion model for galactic
cosmic rays propagation, compatible with B/C analysis  and yielding
the maximal, median and minimal primary antiproton flux \cite{pbar_susy}.
\label{table:prop}}
\end{center}
\end{table}

We proceed now to analyze the extent of compatibility of the neutralino configurations
which fit the DAMA results
 with the present data on cosmic antiprotons. Among the six sets of values for
the parameters $v_0$ and $\rho_0$ analyzed in Figs. \ref{fig:03}--\ref{fig:04},
  let us consider the following ones (the same already considered in Ref. \cite{zooming}):
$A$)  $v_0 = 170$ km sec$^{-1}$, $\rho_0^{\rm min} = 0.20$ GeV cm$^{-3}$;
$B$)  $v_0 = 170$ km sec$^{-1}$, $\rho_0^{\rm max} = 0.42$ GeV cm$^{-3}$;
$C$) $v_0 = 220$ km sec$^{-1}$, $\rho_0^{\rm min} = 0.34$ GeV cm$^{-3}$.

In Figs. \ref{fig:06}--\ref{fig:07} we give the antiproton fluxes
 at $\bar p$ kinetic energy $T_{\bar p}=0.23$ GeV, as
  a function of the neutralino mass  for a cored isothermal halo and
  for the neutralino configurations selected by the DAMA regions shown
  in Fig. \ref{fig:03}--\ref{fig:04}, respectively. Fig. \ref{fig:06}
  refers to the case in which channeling is not included in the derivation of the
  DAMA regions, Fig. \ref{fig:07} to  the case with channeling included.

\begin{figure*}[t] \centering
\vspace{-20pt}
\includegraphics[width=2.0\columnwidth]{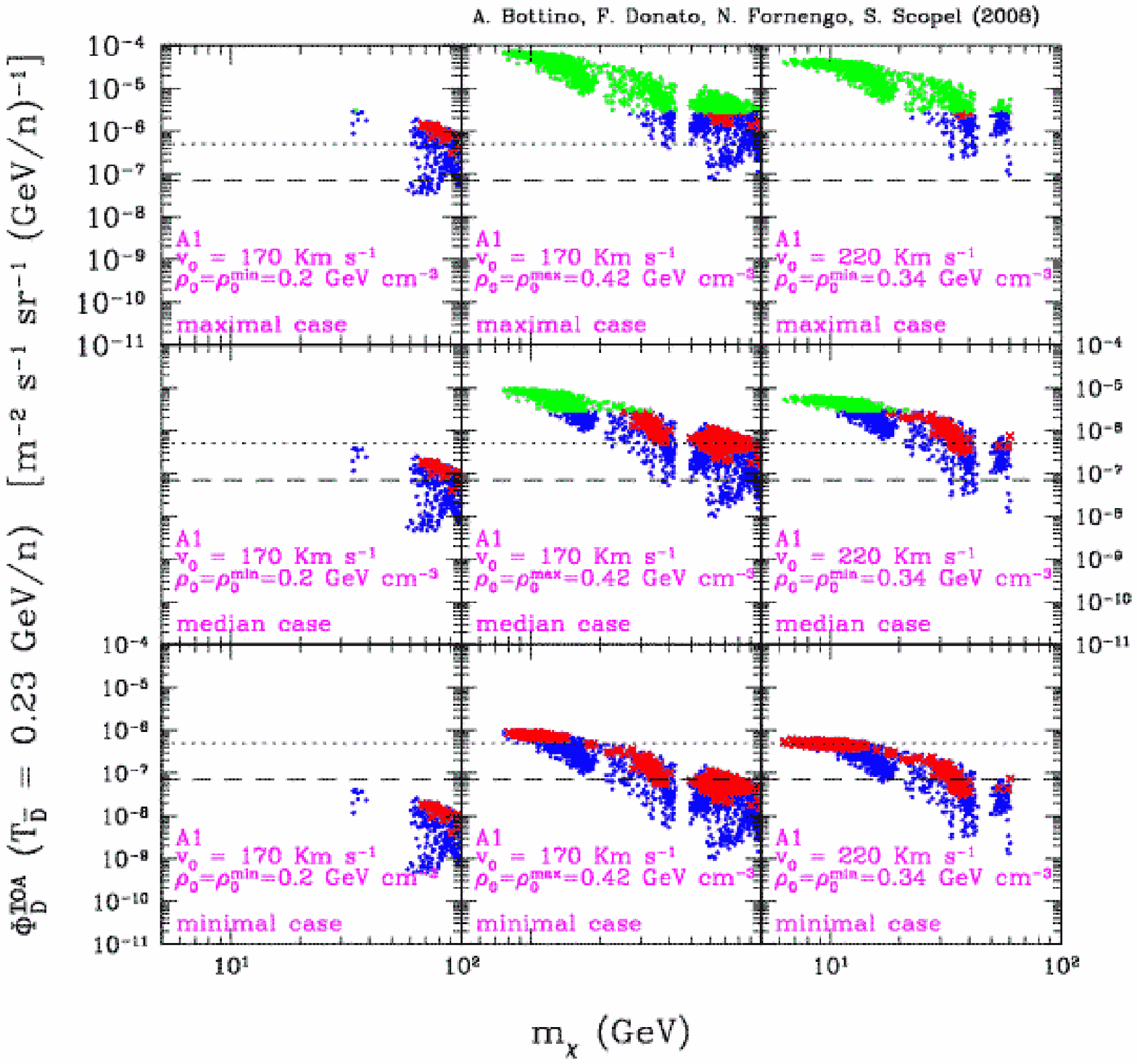}
\vspace{-30pt}
  \caption{Antideuteron flux, at $\bar D$ kinetic energy $T_{\bar D}=0.23$ GeV/n,
    generated by neutralino
  configurations selected by the DAMA data when
a cored isothermal halo is employed and
  the channeling effect is not included.
Each row
  corresponds to a different set of cosmic rays propagation parameters: the
  upper, median and lower rows refer to the sets which provide the maximal,
  median and minimal antiproton flux, respectively (see Table I).
  The three columns refer to the sets of $v_0$ and $\rho_0$ values denoted as sets $A$, $B$ and
$C$ in the text.
  The bold (colored) points refer to configurations
  compatible with the DAMA regions  for the case in which the channeling effect
  is not included (see Fig. \ref{fig:03}). In selecting the allowed configurations the hadronic
  uncertainties have been taken into account. The points are
  differentiated as follows: (red) crosses denote configurations with a
  neutralino relic abundance which matches the WMAP cold dark matter amount
  ($0.098 \leq \Omega_{\chi} h^2 \leq 0.122$), while (blue) dots refer to
  configurations where the neutralino is subdominant ($\Omega_{\chi} h^2 <
  0.098$)
 The light grey (green on-line) points
denote  supersymmetric configurations yielding
an exceedingly high antiproton flux (see Fig. \ref{fig:06}).
The horizontal lines refer to estimated
  sensitivities to antideuterons of the GAPS (dashed) and AMS (dotted) detectors.}
\label{fig:08}
\end{figure*}

\begin{figure*}[t] \centering
\vspace{-20pt}
\includegraphics[width=2.0\columnwidth]{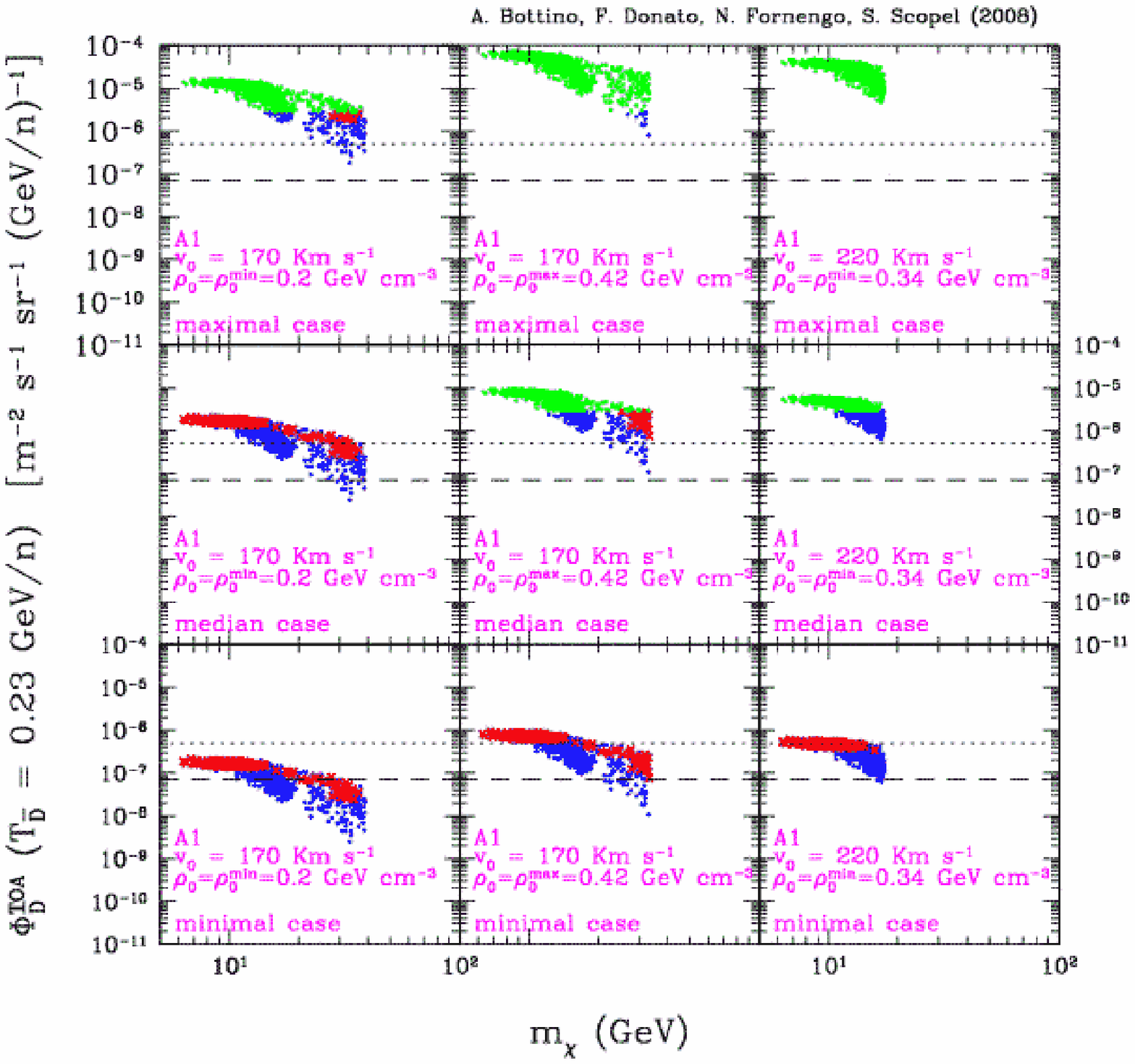}
\vspace{-30pt}
  \caption{Antideuteron flux, at $\bar D$ kinetic energy $T_{\bar D}=0.23$ GeV/n,
   generated by neutralino
  configurations selected by the DAMA data when
a cored isothermal halo is employed and
  the channeling effect is  included (see Fig. \ref{fig:04}).
The light grey (green on-line) points
denote  supersymmetric configurations yielding
an exceedingly high antiproton flux (see Fig. \ref{fig:07}).
Other  notations are
  as in Fig. \ref{fig:08}, except that here
  the neutralino configurations of the scatter plot are those selected by the
  DAMA regions of  Fig. \ref{fig:04}.}
\label{fig:09}
\end{figure*}

At variance with what displayed in Figs. \ref{fig:01}--\ref{fig:05}, where the scatter plot
was evaluated at the reference point: $g_{u,ref}$ = 123 MeV, $g_{d,ref}$ = 290 MeV, in
Figs. \ref{fig:06}--\ref{fig:07} the scatter plots are the results not only of the scan
over the supersymmetric parameter space but also of the variations of the
quantities $g_u$ and $g_d$, as given by Eqs. (\ref{eq:l})--(\ref{eq:g}), when
$\sigma_{\pi N}$ and $\sigma_0$ are varied in the ranges of Eq. (\ref{eq:q}) and
Eq. (\ref{eq:0}), respectively, and $r$ is put at the default value
$r = 25$.

From the results shown in Figs. \ref{fig:06}--\ref{fig:07} we see that, though a number
of configurations are excluded by the BESS data \cite{bess}, many others are perfectly compatible
with BESS and in principle accessible to PAMELA \cite{pamela} and AMS-02 \cite{ams-pbar}.
More specifically, for set $A$, most of the neutralino configurations are unconstrained by the galactic
antiproton data, except for a group of them in the case of the maximal set of the diffusion parameters
and when channeling is included; a sizable number of configurations are at the level of possible
investigation. Sets $B$ and $C$, due to their
corresponding  higher values of $\rho_0$,  are more sensitive to the $\bar p$ constraints
(but also, to a large extent, accessible
to PAMELA and AMS), though prevalently for sets of the diffusion
parameters close to the maximal set.

It is worth noticing explicitly that for the other sets of $v_0$ and $\rho_0$ discussed in
Sect. \ref{sec:regions_mod} and in  Figs. \ref{fig:03}--\ref{fig:04}, but not considered here, one would obtain
plots similar to the ones displayed in Figs. \ref{fig:06}--\ref{fig:07} with scatter plots rescaled
according to the power $\rho_0^2$; thus for these sets the $\bar p$ constraints would be more severe
than in the previous cases.

As we noticed above, a sizable number of neutralino configurations are at the level of the
sensitivities of current experiments on cosmic antimatter. However, for the reasons explained
above, the corresponding primary fluxes would be rather difficult
to be disentangled from the secondary flux (notice that primary and secondary fluxes
have also a very similar behavior as functions of the $\bar p$ kinetic energy) \cite{pbar_susy}.
The cosmic antiproton data, powerful in providing
stringent constraints, are somewhat problematic in providing positive signals for exotic production.
Different is the case for antideuterons which we discuss in the following section.

\begin{figure}[t]
\centering
\vspace{-80pt}
\includegraphics[width=1.15\columnwidth]{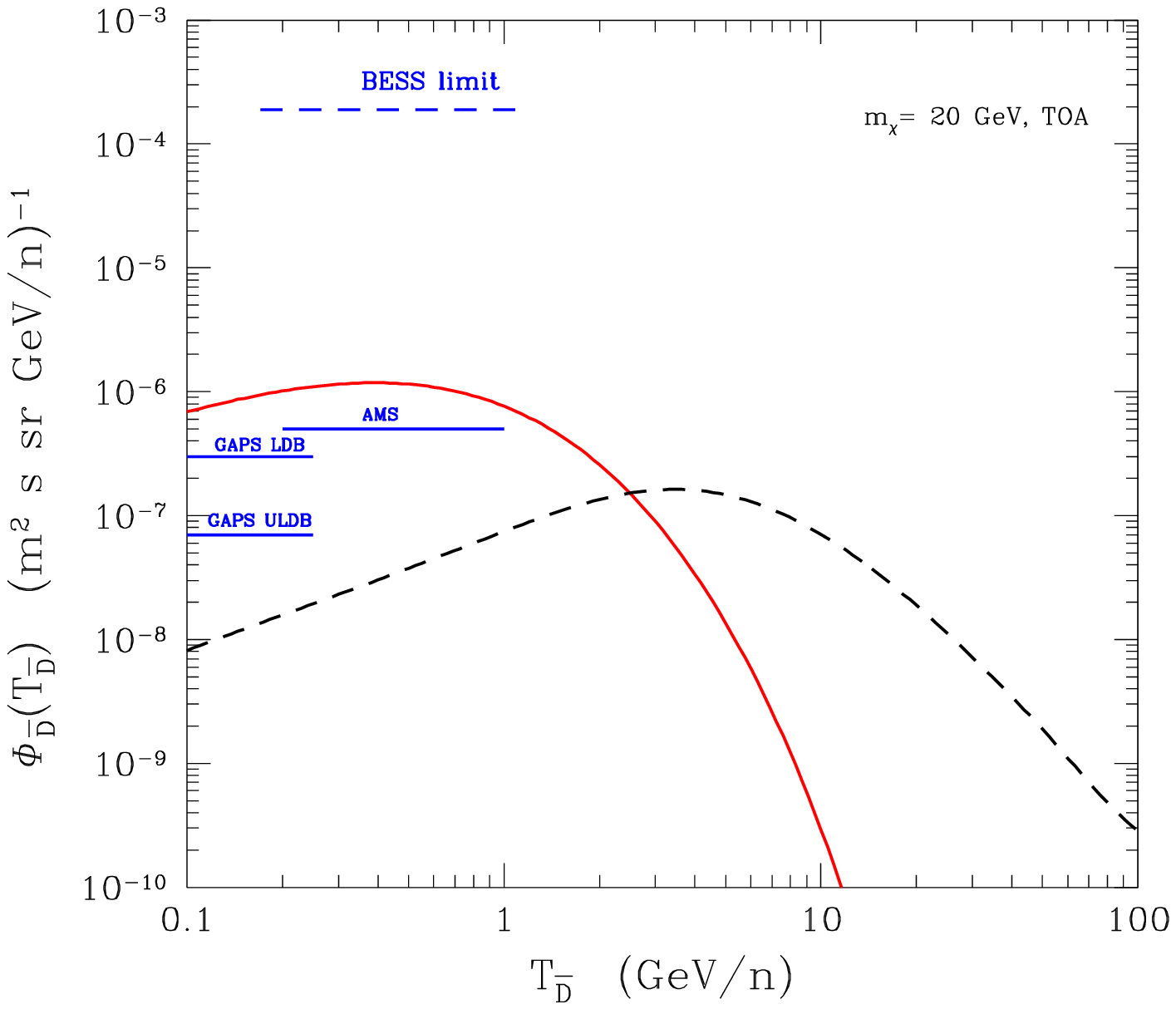}
\vspace{-30pt}
\caption{Solid line: antideuteron flux for a DM halo of $m_\chi=20$ GeV
neutralinos
compatible with the DAMA modulation effect. Dashed line:
secondary component \cite{dbar2008}. Propagation parameters are set at the
median configuration of Table I.
 The upper dashed horizontal line is the
present  BESS upper limit on the search for cosmic antideuterons. The
three horizontal solid (blue) lines are the estimated sensitivities for (from
top to bottom): AMS-02, GAPS on a LDB and GAPS on an ULDB.
}
\label{fig:10}
\end{figure}

\section{Search for antideuterons in the galactic halo}
\label{sec:antideuterons}

In Ref. \cite{dbar} it was shown that the antideuteron spectra derived
from  DM self--annihilation are much flatter than the expected astrophysical
component  for kinetic energies $T_{\bar{D}}\lsim$ 2-3 GeV/n.
Recently,  the calculation of the primary and  secondary antideuteron
fluxes  has been performed \cite{dbar2008}  in the framework of the
full propagation model  outlined in the previous section, encoding all
the possible uncertainty sources.  In Ref. \cite{dbar2008,bp} it has been
shown that antideuterons offer a very promising signal for the
indirect detection  of intermediate and low mass DM particles by means
of future detectors such  as GAPS on long and ultra-long duration balloon
(LDB and ULDB) flights \cite{gaps} and AMS-02 for three years data taking
\cite{ams}.

In the present Section, we present results on
the fluxes of antideuterons produced from DM self--annihilation in the galactic
halo following the procedure explained in Ref.
\cite{dbar2008} (we refer to this paper for all details).
In Fig. 8 we display the antideuteron flux calculated at ${\bar{D}}$ kinetic
energy of $T_{\bar{D}}=0.23$ GeV/n as a function of the neutralino mass
for the cored isothermal halo A1 of Eq. (\ref{isot}). The scatter plot represents the
supersymmetric configurations selected by DAMA data when the channeling
effect is not included, while Fig. 9 shows the same quantities when the
channeling effect is included. The configurations have been selected
with the same method as in the previous Section.
The three upper, median and lower figures correspond to the maximum, median
and minimum set of propagation parameters of Table I. The dotted and dashed
horizontal lines assess the sensitivities of AMS-02 and GAPS ULDB,
respectively.
The light gray (green) points correspond to supersymmetric configurations yielding
an exceedingly high antiproton flux (see Figs. \ref{fig:06}--\ref{fig:07}).
Figs. \ref{fig:08}--\ref{fig:09} show that a sizable number of neutralino
configurations compatible with  the annual modulation data
can generate signals accessible to antideuteron searches planned for the next years. It is notable
that a bunch of configurations not accessible to antiproton signals
could instead be probed by antideuteron searches.

In Fig. \ref{fig:10}, by way of example, we show
the antideuteron flux for a DM halo made of  $m_\chi=20$ GeV
neutralinos compatible with the DAMA effect, with $\rho_0=0.34$ GeV cm$^{-3}$
and setting the propagation parameters to the median values.
The lower, dashed line corresponds to the secondary
antideuteron flux as obtained in Ref. \cite{dbar2008}.
Fluxes are modulated at solar minimum.
The three horizontal lines are the estimated sensitivities for (from
top to bottom): AMS-02, GAPS LDB and GAPS ULDB flights.
The primary flux stands well above the background and exceeds the
experimental reach of next generation detectors.
The result is noticeable: both GAPS and AMS-02 will have the capability
of clearly detecting antideuterons produced from a neutralino halo
compatible with the positive signal found in DAMA data.

\section{Measurements at LHC}
\label{sec:lhc}

Naturally, the viability of the interpretation of the DAMA annual modulation
data in terms of relic neutralinos depends ultimately on the results which LHC will provide
on supersymmetric theories. A thorough investigation on the search
for light neutralinos at LHC has been recently carried out in Ref. \cite{lhc}.
     In this paper two specific scenarios are analysed, both of which dictated by cosmological
properties pertaining to light neutralinos.

A first scenario, scenario $\mathcal{A}$,
 is identified by the following sector of the supersymmetric
parameter space: $M_1 \sim$ 10 GeV, $|\mu| \sim$ (100 - 200) GeV, $m_A
\sim$ 90 GeV, $\tan \beta \sim$ 30 - 45, -1 $\lsim A \lsim$ +1; the
other supersymmetric parameters are not {\it a priori} fixed.
In this scenario the cosmological bound
$\Omega_{\chi} h^2 \leq (\Omega_{CDM} h^2)_{\rm max}$ is satisfied
because of a neutralino self--annihilation mediated by the light
$A$--boson, combined with a high value of $\tan \beta$ and a sizeable
bino-higgsino mixing in the neutralino composition.

A second scenario, {Scenario $\mathcal{B}$},
is identified by the following sector of the supersymmetric
parameter space: $M_1 \sim$ 25 GeV, $|\mu| \gsim$ 500 GeV, $\tan \beta
\lsim$ 20; $m_{\tilde{l}} \gsim$ (100 - 200) GeV, $-2.5 \lsim A \lsim +2.5$;
the other supersymmetric parameters are not {\it a priori} fixed.
In this case the cosmological
constraint $\Omega_{\chi} h^2 \leq (\Omega_{CD M} h^2)_{\rm max}$ is satisfied because of
 stau-exchange contributions
(in the {\it t, u} channels) to neutralino self--annihilation cross section.

In Ref. \cite{lhc} the signals expected at LHC in the two scenarios are discussed
through the main (sequential and branched) chain processes, started by a squark
produced in the initial proton-proton scattering.
Branching ratios and the expected total number of events are
derived in the various benchmarks defined within the two scenarios.

On the basis of these results it is proved in Ref. \cite{lhc}
that LHC should allow an efficient exploration
of the supersymmetric parameter regions compatible
with light neutralinos. Due to the  characteristic features of these regions, the
measurements of LHC  should be able to prove or disprove the supersymmetric model considered in the
present paper and then validate or not our interpretation of the annual
modulation effect in terms of relic neutralinos.

\section{Conclusions}
\label{sec:conclusions}

In this paper we have analyzed the most recent experimental data on
direct searches for dark matter particles in the galactic halo.
We have discussed the various features  involved in an interpretation of these data in terms
of relic neutralinos; the supersymmetric scheme employed is an effective MSSM scheme
 at the electroweak scale, without  gaugino--mass unification at a Grand Unified
scale.
The role of  the uncertainties affecting the neutralino--quark couplings,
because of the involved hadronic quantities, are critically discussed and included
in our considerations.

First we have considered in detail the results of the DAMA Collaboration which, by a
combined analysis of the DAMA/NaI and the DAMA/LIBRA experiments,
for a total exposure of 0.82 ton yr,  provide evidence of an annual modulation
effect at 8.2 $\sigma$ C.L.
Comparison of our theoretical evaluations with the DAMA data has been carried out both in the case
in which channeling effect is included in the DAMA analysis and in the one where it is excluded.
The DAMA results presented here refer uniquely to the case in which the putative WIMP has a
coherent interaction with the nuclei in the detector and are therefore translated in terms of
regions  in the $m_{\chi}$--$\xi \sigma_{\rm scalar}^{(\rm nucleon)}$ plane,
 for different selections of the galactic distribution functions \cite{dama3}.
 We wish also to recall  that
DM--detector interaction mechanisms or DM candidates different from the one considered
in the present paper could be the origin of the annual modulation effect \cite{dama/libra}.

The roles the hadronic uncertainties and of the effect of channeling are
kept separated at any stage of our discussion on DAMA data; the analysis has been
 carried out in such a way that,
 once hopefully some of the afore mentioned
uncertainties are resolved in the future, one can easily employ our present results to narrow down
the physical relevant regions.

By considering first the DAMA data  when a convolution of galactic distribution functions
is considered, we have shown that the annual modulation region is largely compatible
with the theoretical predictions for relic neutralinos with masses
$m_{\chi} \lsim$ 100 GeV, and in particular for neutralinos within the low--energy funnel
for $m_{\chi} \lsim$ 50 GeV. This occurs, whether or not the channeling effect is included.

We have then pursued our analysis by
employing the cored isothermal sphere DF and  shown that
relic neutralinos fit quite well the annual modulation data. For the case where
the channeling is included according to the modeling of
Ref. \cite{channeling}, experimental data favor values
of  $v_0$ and $\rho_0$ which are in the low-medium
side of their own physical ranges, {\it i.e.} $v_0 \simeq$ (170 -- 220) km sec$^{-1}$
and $\rho_0 \simeq (0.3 - 0.4)$ GeV cm$^{-3}$ and neutralino masses
in the mass range $m_{\chi} \simeq (7 - 30)$ GeV.
In case of no-channeling,  low values of $v_0$ and $\rho_0$
($v_0 \simeq$ 170 km sec$^{-1}$ and $\rho_0 \simeq 0.2$ GeV cm$^{-3}$)
appear to be somewhat disfavored, though in this case neutralinos with
$m_{\chi} \simeq$ 60--100 GeV could be involved.  The agreement between
experimental data and theoretical evaluations looks very good
for the other combinations of $v_0$ and $\rho_0$ values,
 with an
overall preference for neutralinos of low mass.
We have also commented about the results one would obtain when some other DF is selected.

The neutralino populations selected on the basis of the annual modulation data have been
analyzed in terms of the antiproton fluxes which they would produce in our halo. We have shown that
many of them are fully compatible with the current stringent bounds on cosmic antiprotons,
especially for values of
local dark matter density
$\rho_0$ and local rotational velocity $v_0$
 in the low side of their physical ranges, and for
values of the diffusion parameters not too close to the values of their maximal set.

It was also derived that forthcoming measurements of galactic antideuterons will have good chances to
explore fluxes generated by the self--annihilation of the neutralino configurations
compatible with the annual modulation data.

The upper bounds on   $\xi \sigma_{\rm scalar}^{(\rm nucleon)}$,  derived by other
experiments of WIMP direct searches \cite{taup2007,xe,cdms,kims}, have  been analyzed
 by considering various WIMP galactic DFs and by including the uncertainties induced by
 the previously mentioned hadronic quantities. It is concluded that no conservative
 upper bound can now exclude the light  ($m_{\chi} \lsim$ 50 GeV) neutralino population within the model discussed in
Sect. \ref{sec:susy}.

We have pointed out
that LHC is expected to  allow an efficient exploration
of the supersymmetric parameter regions compatible
with light neutralinos and the annual modulation data. Due to the characteristic features of the physical regions involved, the
measurements of LHC  could be able to prove or disprove the supersymmetric model considered in the
present paper and then validate or not our interpretation of the annual
modulation effect in terms of relic neutralinos.

It is finally worth noting that the relic neutralinos of light masses, 
considered in the present paper and
previously in Refs. \cite{lowneu,ind}, can be of relevance also in many interesting
astrophysical contexts
different from the ones discussed before. For instance, by their
self--annihilations, these neutralinos
may be the origin of electron--positron pairs which could interact with photons of the
cosmic microwave background producing a peculiar Sunyaev--Zeldovic effect
\cite{cola,cola2}.
Also, positrons and electrons created by annihilation of light neutralinos in the
galactic center
might generate a synchrotron radiation responsible for an excess of microwave
emission from
a region within $\sim 20^o$ of the galactic center \cite{fink} (see also Ref.
\cite{haze}).
Light neutralino dark matter could also imply a sizable primordial $^6$Li abundance
\cite{kj}, a
feature particularly interesting in view of recent new determinations of this abundance
in metal--poor halo stars \cite{asplund}.

\acknowledgments
We thank the DAMA Collaboration for providing us with the
results of their analysis prior to their publication.
 We are also grateful to Rita Bernabei and
 Pierluigi Belli for useful discussions.
 We  acknowledge Research Grants funded jointly by Ministero dell'Istruzione,
dell'Universit\`a e della Ricerca, by Universit\`a di Torino and by
Istituto Nazionale di Fisica Nucleare within the {\sl Astroparticle Physics
Project}.


\begin{thebibliography}{99}
\bibitem{note1} In deriving the regions of Figs. 1-2, cases A, B, C defined
in Sect. 7.2 of the first paper of Ref. \cite{dama/nai} have been included.
\end{thebibliography}
\end{document}